\RequirePackage{fix-cm}

\makeatletter
\def\@cons#1#2{\begingroup\let\@elt\relax\xdef#1{\ifx#1\relax\else#1\fi\@elt #2}\endgroup}
\makeatother

\documentclass[smallextended]{svjour3}       

\usepackage{amsmath,amssymb,amsfonts}
\usepackage{algorithmic}
\usepackage{graphicx}
\usepackage{textcomp}
\usepackage{url}

\usepackage{wasysym}            
\usepackage{natbib}    
\usepackage{hyperref}           
\usepackage{cleveref}           
\usepackage{booktabs}           
\usepackage{tabularx}           
\usepackage[dvipsnames]{xcolor} 
\usepackage{tikz}
\usepackage[all]{nowidow}       
\usepackage[draft]{changes} 
\usepackage{rotating}
\usepackage{enumitem}
\usepackage{multirow}
\usepackage{soul}

\usepackage[utf8]{inputenc}  
\usepackage[T1]{fontenc}     
\usepackage{eurosym}

\renewcommand{\added}[1]{#1}
\renewcommand{\deleted}[1]{}

\newcolumntype{Y}{>{\raggedright\arraybackslash}X}

\newcounter{studiescounter}
\makeatletter
\newcommand{\labelText}[2]{%
#1\refstepcounter{studiescounter}%
\immediate\write\@auxout{%
  \string\newlabel{#2}{{1}{\thepage}{{\unexpanded{#1}}}{studiescounter.\number\value{studiescounter}}{}}%
}%
}
\makeatother

\definecolor{DustySapphire25W}  {HTML}{C3D6E9}
\definecolor{SmokyLavender25W}  {HTML}{E3D8E6}

\newcommand{\dimension}[1]{\sethlcolor{SmokyLavender25W}\hl{#1}}
\newcommand{\theme}[1]{\sethlcolor{DustySapphire25W}\hl{#1}}

\usepackage{draftwatermark}
\SetWatermarkText{PREPRINT}
\SetWatermarkScale{1}

\newcommand{\mypar}[1]{\medskip\noindent\textbf{\textit{#1:}}\ }

\newlist{rqlist}{itemize}{2}
\setlist[rqlist]{nolistsep,leftmargin=1.55cm,rightmargin=0.5cm}

\newcounter{rq}  
\renewcommand{\therq}{RQ-\arabic{rq}} 

\newcommand{\myrq}[1]{%
  \setcounter{rq}{\numexpr#1-1\relax}%
  \refstepcounter{rq}%
  \textbf{\therq}%
  \label{rq#1}
}

\newcommand{\rqanswer}[2]{%
    \fbox{\begin{minipage}{=0.9\linewidth}
      \textbf{\ref{#1} Answer:} #2
    \end{minipage}}
}

\begin{document}


\title{An Explanation of Software Architecture Explanations}

\author{Satrio A. Rukmono \and Filip Zamfirov  \and Lina Ochoa \and Floris Pex \and Michel R.~V. Chaudron}

\institute{
    Satrio A. Rukmono \at
	Eindhoven University of Technology, Netherlands \\
    \email{s.a.rukmono@tue.nl}
    \and
    Filip Zamfirov \at
	Eindhoven University of Technology, Netherlands \\
    \email{f.zamfirov@tue.nl}
    \and
    Lina Ochoa \at
	Eindhoven University of Technology, Netherlands \\
    \email{l.m.ochoa.venegas@tue.nl}
    \and
    Floris Pex \at
	Independent Researcher, Netherlands \\
    \and
    Michel R.~V. Chaudron \at
	Eindhoven University of Technology, Netherlands \\
    \email{m.r.v.chaudron@tue.nl}
}

\date{Received: date / Accepted: date}

\maketitle

\begin{abstract}
Software architecture knowledge transfer is essential for software development, but related documentation is often incomplete or ambiguous, making oral explanations a common means.
Our broader aim is to explore how such explanations might be better supported and eventually automated; as a prerequisite, we first investigate how explanations are actually conducted in practice across five areas: explanation topics, explanation plans, supporting artefacts, typical questions, and expectations and challenges.
We report on semi-structured interviews with 17 software professionals across diverse organisations and countries.
Our findings include that explanations must balance problem- and technical-domain while considering the explainee's role, experience, and goals.
Moreover, explainees frequently seek not only structure and behaviour, but also decision rationale.
We propose the Explanation Window, a framework for focusing information by adjusting functionality scope and detail, and emphasise the importance of including system context. These findings provide an empirical basis for improving architecture explanations and guiding future work on tool support and automation.
\end{abstract}

\section*{Declarations}
\noindent
\paragraph{Competing interest.}
We have no conflicts of interest with any individuals or organisations that could negatively bias our work. 

\noindent
\paragraph{Ethics approval.}
The presented research methodology was evaluated and approved by the Ethical Review Board at Eindhoven University of Technology before conducting the study. A consent form was shared with and signed by the study's participants before the interview to guarantee an ethical approach.

\noindent
\paragraph{Data, materials and/or code availability.}
Data accompanying this article is currently published in Zenodo at \url{https://bit.ly/41XMFi5}.\footnote{The current Zenodo bundle is marked as a draft for review. It will be published once we have obtained the publication approval.}

\noindent
\paragraph{Generative AI use.}
We declare that AI solutions (i.e., Grammarly, Writefull, and ChatGPT) were used for grammar checking and rephrasing. 
We declare that the content of this article and the whole research procedure were developed and fully conducted by the authors, taking full responsibility for the publication's content.

\section{Introduction}\label{sec:intro}

\added{The effective transfer of software architectural knowledge is a critical and challenging aspect of large-scale software development and maintenance~\citep{krancher2020knowledge, nidhra2013knowledge}. 
\textit{Software architecture} refers to the high-level structuring of a system, including its components, their relationships, and the rationale that guides design decisions~\citep{bass2021software,lethbridge2003software}.
In practice, architectural knowledge is commonly transferred through documentation or informal explanations~\citep{robillard2017demand}.
However, architectural documentation is often incomplete, ambiguous, obsolete, or fragmented~\citep{aghajani2019software,ernst2023documentation, uddin2015how} impacting onboarding processes and knowledge continuity~\citep{treude2020beyond, steinmacher2016overcoming, canfora2012who, schroter2017comprehending, ernst2023documentation}.
Meanwhile, there is evidence showing that informal and interactive explanations by architects or experienced developers are an effective alternative to transferring knowledge to colleagues and newcomers~\citep{canfora2012who,dagenais2010moving, unphon2010software}.
We refer to the process by which an architect or experienced developer conveys the system's structures, behaviours, and rationale to a less-informed audience as \textit{software architecture explanation}.}

\added{Despite the prevalence of these software architecture explanations in practice, there is a lack of empirical research on the topic.
To our knowledge, the closest contribution to understanding these explanations comes from the vision of \citet{robillard2017demand}, which advocates an on-demand developer documentation paradigm.
Although the vision uses the term \textit{documentation}, it moves away from traditional documentation tied to a snapshot of a system to discovering and summarising relevant information tailored to the developer needs, similar to how experts explain less-informed audiences.
As a vision, no underlying research is provided, and, therefore, there is still a limited systematic understanding of how software architecture explanations are actually conducted, what topics and artefacts are involved, what questions are typically asked, and what are recurring challenges and expectations from IT professionals.}

\added{Gaining further understanding on software architecture explanations would help comprehending and shaping what constitutes an effective explanation.
Additionally, it could set the ground for automation as one-to-one explanations with experts are an expensive but effective process to maintain software quality, especially in the presence of repetitive turnovers and onboardings~\cite{foucault2015impact}. 
Our findings can then serve as input for future work that will derive criteria to guide the design and evaluation of automated software architecture explanations.}
This study therefore takes the empirical first step of characterising how software architecture explanations take place in real industrial contexts. 
For that, we consider the following research questions:


\begin{rqlist}
    \item[\myrq{1}] \textbf{What key \textit{topics} are addressed in software architecture explanations?} This question examines the specific content discussed, from technical elements (e.g., architectural patterns, views) to contextual issues practitioners prioritise during software architecture explanations.

    \item[\myrq{2}] \textbf{What does a structured \textit{plan} for explaining software architecture entail?} This question explores the diverse activities that typically take place during explanations and their relations.

    \item[\myrq{3}] \textbf{What \textit{supporting artefacts} are used during software architecture explanation?} This question identifies existing or on-the-fly assets such as diagrams and tools. It helps understanding which and how visual and textual aids support comprehension, thereby guiding their refinement to better support architectural explanations.

    \item[\myrq{4}] \textbf{What are the \textit{typical questions} asked during software architecture explanation?} This question examines recurring inquiries from explainees and reveals which aspects of the system need to be elaborated, helping explainers anticipate and address these issues.

    \item[\myrq{5}] \textbf{What are \textit{expectations} and \textit{challenges} during and after software architecture explanations?} This question investigates both what participants expect and the challenges they face to uncover gaps between communication and learning, guiding improvements in knowledge transfer.
\end{rqlist}

We systematically investigate our research questions through a \added{judgment study~\citep{stol2020guidelines} in the form of semi-structured interviews} with practising software professionals. 
We interviewed 17 participants occupying either \textit{explainer} or \textit{explainee} roles.
To ensure a robust and ethically sound investigation, we follow \citeauthor{strandberg2019ethical}'s guidelines~\citeyearpar{strandberg2019ethical}. 
The interview data was analysed using the Gioia methodology~\citep{gioia2013seeking}, combining open and in-vivo coding~\citep{corbin2014basics,manning2017invivo} to identify themes relevant to the practice of software architecture explanation.

\added{Our study shows that: (1) \textit{Topics.} Software architecture explanations revolve around system structures, rationale, and evolution, yet quality attributes remain under-communicated;
(2) \textit{Plan.} Effective explanations integrate diverse activities, such as reviewing documentation, exploring artefacts, asking questions, and engaging with feedback, favouring top-down comprehension and incremental, interactive, in-situ explanation styles;
(3) \textit{Supporting artefacts.} Explanations are anchored in both architecture- and implementation-focused documentation. Structural and behavioural diagrams dominate, with growing use of high-level and data views, while development tools are largely avoided;
(4) \textit{Typical questions.} Explanations address both exploratory questions (probing rationale, impacts, and alternatives) and confirmatory questions (validating partial knowledge) across levels of abstraction and architectural viewpoints; and
(5) \textit{Expectations and challenges.} Explainers and explainees share expectations of clarity, structure, and actionable closure, but face challenges in managing abstraction levels, curating relevant content, and navigating documentation gaps.}
We overlay the theoretical framing of the Twin Peaks model on top of our findings on software architecture explanations to emphasise the importance of separating the ``how'' and the ``what'' and deliberately bridge the (often implicit or presumed) business context and technical rationale. We also introduce the Explanation Window framework as a means of tailoring the scope and depth of architectural explanations to audience needs.

The remainder of this paper is organised as follows. 
\Cref{sec:rw} reviews related work. \Cref{sec:methodology} details our study methodology, including participant selection, data collection, and analysis procedures. \Cref{sec:results} presents our findings, which are then discussed in \Cref{sec:discussion}. \Cref{sec:ttv} examines the trustworthiness of our qualitative study and, finally, \Cref{sec:conclusions} concludes with future research directions.

\section{Related Work}
\label{sec:rw}

Despite the significance of software architecture explanation for effective development and onboarding, there is a remarkable absence of empirical research examining how such explanations are actually conducted in practice. Prior studies predominantly focus on the effectiveness of documentation formats~\citep{ernst2023documentation, shahin2014architectural}, the role of specific artefacts or tools~\citep{hoquang2020interactive, jolak2020software}, or how to effectively perform knowledge transfer~\citep{camacho2013understanding, canfora2012who}, but stop short of analysing the explanation process itself.

For instance, \citet{ernst2023documentation} found that the document format (narrative vs. structured) does not significantly affect newcomers' understanding; prior code exposure and architectural information needs matter more. Similarly, \citet{shahin2014architectural} showed Architectural Design Decisions (ADDs) enhance comprehension but not necessarily task completion time.

The persistent gap between documentation and implementation has led to studies on the informal mechanisms by which architectural knowledge is shared. \citet{unphon2010software} highlight the role of a ``walking architecture'' (through chief architects or lead developers) in managing design and communication beyond documentation.
Related work on tool support, such as RoleViz by \citet{hoquang2020interactive} and graphical design description tools as explored by \citet{jolak2020software}, show that visual artefacts improve usability, communication, and productivity, but textual descriptions can lower perceived explanation quality.

However, knowledge transfer is not simply about accessing artefacts. \citet{camacho2013understanding} stated that knowledge transfer in software engineering requires effective application, not just dissemination. \citet{canfora2012who} noted that informal, face-to-face interactions are the primary source of architectural knowledge for newcomers. This dynamic supports the explainer/explainee distinction and aligns with cognitive apprenticeship models.

Notably, several studies have explored how programmers ask questions when seeking information~\citep{sillito2006questions,kubelka2019live,ratanotayanon2006programmers}, but these focus on immediate, task-specific information needs. In contrast, architectural explanations are conceptual, aiming to build a deep understanding of system design, including decision rationales and component relationships.
While these studies examine developers' question-asking, architectural information transfer differs fundamentally.
Below, we contrast programming questions with architecture comprehension as our initial framing and assumption:

\mypar{Technical task orientation vs. conceptual understanding} Programming aims for immediate and specific solutions (e.g., code snippets), often found in documentation or forums. Architectural explanation is conceptual, seeking deep understanding of system design, decision rationales, and component relationships.

\mypar{Detailed vs. broad information} Programming tasks need immediate and detailed data focused on syntax or functions. Architectural understanding covers broader topics like design patterns, scalability, and trade-offs, building a mental model of the whole system.

\mypar{Exploration vs. transfer} Programming often involves trial and error. Architecture knowledge transfer applies prior expertise to develop lasting insight into complex system interactions.\\

As \citet{arab2022exploratory} suggest, there is a fundamental difference between code sharing and strategic knowledge sharing. Existing work leaves open the question of how architectural explanations are structured, what topics and artefacts are prioritised, how explainer and explainee roles interact, and where the main challenges and breakdowns occur in practice. This study directly addresses that gap. We offer the first open-ended empirical analysis of software architecture explanation as a situated practice, examining not only what information is conveyed, but \emph{how}, \emph{why}, and with what artefacts, strategies, and outcomes. In doing so, we move beyond artefact-centric or tool-focused perspectives, and shed light on the actual communicative work that sustains architectural knowledge in software organisations.

\section{Methodology}
\label{sec:methodology}

\added{We carried out a \textit{judgement study}~\citep{stol2020guidelines} with the primary aim to collect accounts of software architecture explanation practices as they occur in professional environments. To do so, we conducted semi-structured interviews with practitioners who reflected on recent authentic experiences in their own work contexts.}

\added{We initially sought to conduct a \textit{field study} via in-situ observations of architecture explanation sessions; however, all but one organisation we approached declined participation due to confidentiality concerns regarding their software. As a result, we relied on interviews, following best practices for qualitative research when direct access to field settings is restricted~\citep{seaman1999qualitative,stol2020guidelines}. The ABC framework~\citep{stol2020guidelines} recognises that, when observational access is impractical, interviews grounded in lived experience are an accepted alternative to capture the practitioner's context, although they necessarily involve trade-offs in realism and depth.}


\added{As judgement study, i.e., in the absence of direct observation or artefact collection, our data rely on participants' self-reported experiences. In particular, participants were explicitly prompted to anchor their accounts in actual projects and situations, and demographic and background questions were used to contextualise each account. As such, our approach provides context-rich practitioner perspectives, but does not offer the maximal contextual realism or behavioural evidence of immersive fieldwork. We therefore do not claim ethnographic depth, but rather present our results as derived from \textit{contextually rich and practitioner-anchored interviews}, acknowledging both their authenticity and their inherent limitations. This is in line with interpretive qualitative traditions in empirical software engineering, where practical restrictions prevent direct access to field sites~\citep{easterbrook2008selecting}. Researchers and readers should interpret our findings with these constraints in mind. We discuss the limitations of this approach in detail in Section~\ref{sec:ttv}.}

\begin{figure*}[ht]
    \centering
    \includegraphics[width=\linewidth]{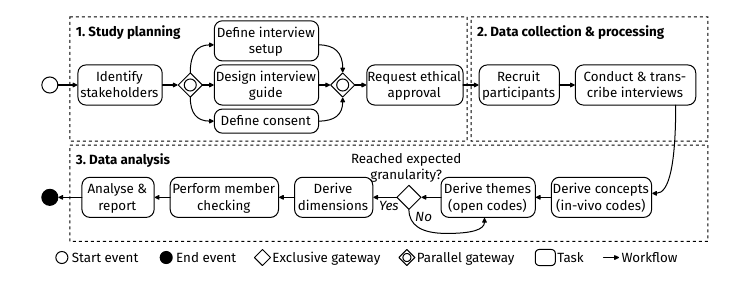}
    \caption{Interview study methodology.}
    \label{fig:methodology}
\end{figure*}

\Cref{fig:methodology} outlines our methodology in three phases: (1) study planning---stakeholder identification, questionnaire development, ethical considerations; (2) data collection---conducting and transcribing interviews; and (3) data analysis---coding, member checking, and reporting.

\subsection{Study Planning}
\label{sec:planning}

The planning phase aims to develop a rigorous and ethical study design.

\mypar{Identify stakeholders} 
\label{sec:stakeholders}
\textit{Stakeholders} in our study context include all people involved in giving or receiving software architecture explanations, as well as their project supervisors or managers~\citep{strandberg2019ethical}. However, only two groups served as study \textit{participants}: \emph{explainers}, typically architects, senior developers, technical leads, or solution architects with deep system knowledge; and \emph{explainees}, who lack familiarity with (parts of) the system. We classified participants by their predominant role in the context of the system under discussion---rather than general professional seniority---to ensure focused, context-specific feedback.

\added{We used a combination of convenience, purposive, and snowball sampling to recruit participants who recently engaged in a software architecture explanation in real-world contexts as either explainers or explainees. Convenience sampling allowed efficient recruitment from accessible professional networks. Purposive sampling targeted professionals with relevant first-hand experience, while snowball sampling extended reach to under-represented or specialised roles. These approaches are widely accepted in empirical software engineering when depth and context matter~\citep{sedgwick2013convenience,tongco2007purposive,edgar2017exploratory,seaman1999qualitative,easterbrook2008selecting,stol2016grounded}.
\\
We did not use general online platforms such as Prolific, as these are designed for studies seeking a broad demographic reach and cannot guarantee access to professionals embedded in specific, confidential organisational settings~\citep{stol2016grounded,reid2022software}. This does not align with the objectives of a judgement study, which requires careful selection of experts~\citep{stol2020guidelines}. Previous studies~\citep{reid2022software,ebert2022recruiting} also indicated that such platforms risk data noise and low trustworthiness when used for context-dependent topics. With the increasing availability of generative AI, ensuring the authenticity and quality of responses on such platforms has become even more challenging. 
\\
Given the small number of participants, we do not claim statistical generalisability. This study is positioned as a qualitative study, aimed at an insight into professional practices rather than broad representativeness.}

\begin{sloppypar}
\mypar{Define interview setup}
Interviews were conducted in English and held online using Microsoft Teams. Only the interviewer and interviewee were present (except for the first interview, which included a silent observer for interviewer feedback). Microsoft Teams was chosen for secure recording and GDPR-compliant transcript generation. All personal and confidential data were anonymised by the first author before being shared with co-researchers to avoid leaving traces that could identify participants. Conducting interviews in a private setting helped reduce distractions~\citep{stol2020guidelines}, bias, or retaliation risk~\citep{strandberg2019ethical}.
\end{sloppypar}

\mypar{Design interview guide}
\added{The interview guide design is connected to the initial plan for in-situ observation. As mentioned earlier, we succeeded in observing a real-world knowledge transfer session at one company, where a departing architect ($A$) explained system architecture to another architect from a different team ($B$), who was going to assume temporary responsibility for the project. This observation informed the formulation of our research questions and thus the drafting of the interview guide.}

\added{Subsequently, we conducted a pilot interview with a newly hired architect ($C$), who later participated in an onboarding session with $B$. Following this, $B$, having both received an explanation and acted as explainer, provided detailed feedback on the interview process and questions. This dual-role perspective was instrumental in refining the guide.  The pilot interview served as a realistic chance to enhance our interview design prior to conducting the full study~\citep{linaker2015guidelines,strandberg2019ethical}. Data from the pilot were excluded from the main analysis.}

The resulting interview guide was structured as a flexible script~\citep{bird2016interviews,smulowitz2017interview}, with separate versions for explainers and explainees. The guide began with instructions to ground responses in a real-work context, i.e., a specific architecture explanation the participant had experienced, not limited to face-to-face situations. 
It was followed by demographic questions, which cover experience, company, project, role, tenure, and context, and were used to assess participant suitability for their assigned role.
We included core questions targeting the five dimensions aligned with our research questions: explanation topics (\ref{rq1}), explanation plans (\ref{rq2}), supporting artefacts (\ref{rq3}), typical questions (\ref{rq4}), and expectations and challenges (\ref{rq5}). We combined specific and open questions to gather in-depth and nuanced views~\citep{hove2005experiences,seaman1999qualitative}.
These questions and their mapping to the study's research questions are listed in \Cref{tab:questions}.
To control for quality, two of the authors drafted the guide, which was then evaluated and further refined with input from a software architecture expert and the pilot interview mentioned before.

\begin{table*}[t]
\centering
\footnotesize\sffamily
\caption{Interview questions specific to explainers and explainees.}
\label{tab:questions}
\begin{tabularx}{\textwidth}{@{}lYc@{}}
\toprule
\multicolumn{1}{l}{\textbf{ID}} & \multicolumn{1}{c}{\textbf{Question}} & \multicolumn{1}{c}{\textbf{RQ}} \\ \midrule
\multicolumn{3}{c}{\textbf{For Explainers}} \\ \midrule
$Q_{1_E}$ & What needs to be explained about a system’s architecture? & \ref{rq1} \\
$Q_{2_E}$ & What are important considerations you keep in mind when explaining a software system? & \ref{rq1}, \ref{rq2} \\
$Q_{3_E}$ & How do you start an architecture explanation? & \ref{rq2}  \\
$Q_{4_E}$ & Can you elaborate on the whole explanation process? &  \ref{rq2} \\
$Q_{5_E}$ & How do you wrap up the explanation? & \ref{rq2} \\
$Q_{6_E}$ & Do you use any tools, artifacts, or frameworks at any point in your explanation? & \ref{rq3} \\
$Q_{7_E}$ & Could you describe the typical questions you receive from team members during architecture explanations? & \ref{rq4} \\
$Q_{8_E}$ & Have you encountered any challenging or unexpected questions during architecture explanations? How did you handle them? & \ref{rq4} \\
$Q_{9_E}$ & What are your expectations after explaining an architecture to the other software professional within your project? & \ref{rq5} \\ \midrule
\multicolumn{3}{c}{\textbf{For Explainees}} \\ \midrule
$Q_{1_e}$ & What do you expect from a software architecture explanation? & \ref{rq5} \\
$Q_{2_e}$ & What makes a good software architecture explanation? & \ref{rq1}, \ref{rq2} \\
$Q_{3_e}$ & What level of detail would you prefer an architecture explanation to have? & \ref{rq1} \\
$Q_{4_e}$ & What considerations are needed to comprehend the architecture of a software system? & \ref{rq1}, \ref{rq2} \\
$Q_{5_e}$ & How do you start approaching (to understand) the architecture of a software system? & \ref{rq2} \\
$Q_{6_e}$ & What other aspects do you consider important in that process? & \ref{rq1}, \ref{rq2} \\
$Q_{7_e}$ & What do you expect to get at the end of a software architecture comprehension task? & \ref{rq5} \\
$Q_{8_e}$ & Do you use any tools or artifacts at any point when trying to understand the architecture of a system? & \ref{rq3} \\
$Q_{9_e}$ & Could you describe the type of relevant questions you ask software architects during architecture explanations? & \ref{rq4} \\
$Q_{10_e}$ & Have you encountered any difficulties in understanding architecture explanations? If so, what were they, and how did you address them? & \ref{rq1}, \ref{rq2}, \ref{rq3} \\
$Q_{11_e}$ & Are there any specific types of information or details that you find particularly useful in software architecture explanations? & \ref{rq1} \\ \bottomrule
\end{tabularx}
\end{table*}

\mypar{Define consent}
We follow \citet{kitchenham2015evidence} for ethical considerations. Ethical approval was granted by the Ethics Review Board (ERB) of Eindhoven University of Technology on 10 July 2023, ensuring compliance with all relevant standards, laws, and regulations. All participants signed informed consent before participation; they were told participation was voluntary and could be withdrawn at any time.

\subsection{Data Collection \& Processing}

\mypar{Recruit participants}
We invited 20 professionals: 10 explainers and 10 explainees. Nine explainers and eight explainees accepted, yielding acceptance rates of 90\,\% and 80\,\%, respectively. Prior work suggests that thematic saturation is typically reached after 12--15 interviews~\citep{guest2006many,guest2020simple}. \added{Participant characteristics are summarised in \Cref{tab:participants}. They represent diverse cultural and professional backgrounds and are based in the Netherlands, Sweden, Germany, the United States, Argentina, and Colombia. While individual employers cannot be disclosed, \Cref{tab:turnovers} reports annual revenue ranges of the organisations where participants work, based on publicly available financial information. This shows that most participants are employed in large multinational corporations, underscoring that our study covers explanations in real, large-scale industrial projects.}
\added{Moreover, explainers typically have 8+ years of relevant experience. In one micro-enterprise case, the explainer had three years' tenure, which is the norm in that organisation. In another case, an explainer has only one year of experience at the current company but with many years of experience as an architect in a similar domain and system type. Conversely, some explainees have extensive professional experience, but they were all consistently less familiar with the system being explained in their own individual circumstances.}

\begin{sidewaystable}
\vspace*{0.44\textheight}
\centering
\footnotesize\sffamily
\caption{Interview study participants. \textit{Comp. stands for company and Proj. for project.}}
\label{tab:participants}
\begin{tabularx}{\textwidth}{@{}clYYYll@{}}
\toprule
\textbf{ID} & \textbf{Experience} & \textbf{Comp. type} & \textbf{Proj. type} & \textbf{Role} & \textbf{Comp. time} & \textbf{Proj. time} \\ \midrule
\multicolumn{7}{c}{\textbf{Explainers}} \\ \midrule
$E_{1}$ & 15 years & Equipment manufacturer & Metrology & Software architect & 11 years & N/A \\ 
$E_{2}$ & 15 years & University & Information system & Software architect & N/A & 1.5 years \\ 
$E_{3}$ & 14 years & Location technology & Microservices & Senior software engineer & 4 years & 2--3 years \\ 
$E_{4}$ & 15 years & Technology & Various & Solutions architect & 2 years & 2 years \\ 
$E_{5}$ & 19 years & Technology & Consultancy & Solutions architect & 12 years & N/A \\ 
$E_{6}$ & 25 years & Software engineering & Embedded system for medical device & Software architect & $>$1 year & 1 year \\ 
$E_{7}$ & 3 years & Software engineering & POS system & Technical lead & 3 years & 3 years \\ 
$E_{8}$ & 8 years & Transportation products and services & Search engine and automotive & Software/ System architect & 4 years & 1.5 years \\ 
$E_{9}$ & 21--22 years & Research organisation & Metaprogramming toolkit & Consultant & 24 years & 20 years \\ \midrule 
\multicolumn{7}{c}{\textbf{Explainees}} \\ \midrule
$e_{1}$ & 8 years & Consultancy & Data analytics & Data analytics specialist & 5 months & 3 months \\ 
$e_{2}$ & 1.5 months & Research organisation & Maintenance of legacy system & Researcher & 1.5 months & 1.5 months \\ 
$e_{3}$ & 2 months & Applied research & Formal methods in industry & Researcher & 2 months & 2 months \\ 
$e_{4}$ & 2 months & Health technology & Embedded system for medical device \deleted{(?)} & Intern & 2 months & 2 months \\ 
$e_{5}$ & 35 years & Consultancy & Various & Architecture reviewer & 30 years & 1--3 years \\ 
$e_{6}$ & 2 months & Vehicle manufacturer & Real-time embedded system & Functional application manager, user acceptance tester & 2 months & 3--4 months \\ 
$e_{7}$ & 9 years & Research agency & Embedded systems & Technologist, system architect, system engineer & 5 months & 5 months \\ 
$e_{8}$ & 4.5 years & Equipment manufacturer & Refactoring legacy code & Intern & 6 months & 6 months \\ \bottomrule 
\end{tabularx}
\end{sidewaystable}

\begin{table}[h]
    \centering
    \footnotesize\sffamily
    \caption{\added{Number of study participants by employer's annual revenue and organisation size.}}
    \label{tab:turnovers}
    \begin{tabularx}{.8\linewidth}{YYl}
        \toprule
        \textbf{Number of participants} & \textbf{Employer's recent annual revenue} & \textbf{Organisation category}\\ \midrule
        1 & (N/A) & Freelance (client not disclosed)\\
        3 & (N/A) & Academic organisation\\
        1 & (N/A) & Micro enterprise*\\
        3 & \euro 10--50 million & Medium enterprise*\\
        2 & \euro 500--700 million & Large enterprise*\\
        5 & \euro 10--50 billion & Large enterprise*\\
        2 & \euro 200--700 billion & Large enterprise*\\
        \bottomrule
    \end{tabularx}
  \vspace{0.5em}
  
\rmfamily
*European Commission classification \citep{ec2003sme}.
\end{table}

Onboarding was expected as the main scenario for explanation, but only four participants reported it; other scenarios included research, \added{where a data scientist examines the team's software to identify data schemas and flows}; solution-provider-to-client explanation, \added{where a solutions architect tries to match their solutions with the client's requirements}; cross-team collaboration, \added{where different companies are involved in engineering an (eco)system}; and architecture evaluation and audit, \added{where a team member must explain the architecture to an auditor}. These were identified from demographic data and thus not included as findings.

\mypar{Conduct \& transcribe interviews}
The first author conducted all interviews between August 2023 and January 2024.
Interviews were transcribed using automatic speech recognition built into Microsoft Teams, followed by manual proofreading and anonymisation (removal of names, company references, etc.).

\subsection{Data Analysis}

We adopted the Gioia methodology~\citep{gioia2013seeking} for its strengths in providing transparency and rigour in inductive qualitative analysis. In this context, \textbf{concepts} refer to meaning-laden units drawn directly from participant statements (in-vivo codes), \textbf{themes} represent recurring patterns that capture shared ideas or concerns, and \textbf{dimensions} are broader, higher-level categories that organise and relate themes.

Our analysis relied on a common foundational assumption in interview-based research that participants can reflect on and articulate their experiences. ATLAS.ti\footnote{\label{ref:atlas.ti}\url{https://atlasti.com/}} software was used to manage and facilitate the coding process.

\mypar{Derive concepts}
First-order concepts were identified using in-vivo coding~\citep{manning2017invivo} by the first four authors. \added{The first three authors split the interview transcripts evenly among themselves and analysed the complete transcripts. Separately, the fourth author analysed all transcripts, but focussing solely on parts related to questions being asked during explanations (i.e., for \ref{rq4}).}

\mypar{Derive themes}
Second-order themes were developed through open coding~\citep{corbin2014basics}, with the second and third authors independently coding the same transcripts in iterative rounds. Each round covers four transcripts. After each round, the coders met to discuss and align their codes, documenting the rationale for additions or changes in individual codebooks. Each author maintains an individual codebook to record any additions or changes made during coding. These records include notes on the rationale for each change. During alignment meetings, overlapping or redundant codes are identified, merged, and updated in both the shared ATLAS.ti project and a consolidated codebook. All agreed-upon code additions, removals, or modifications are systematically documented in the shared codebook to track the development of the coding scheme across interviews.

Although the Gioia approach discourages imposing theoretical constructs prematurely, we pragmatically used Kruchten's 4+1 view model~\citep{kruchten19954+1} and comprehension models~\citep{alsaiyd2017source,storey2005theories} as \added{pre-existing thematic categories}. While these frameworks provided analytic structure and terminology, they did not constrain inductive openness; any codes and themes that fell outside these frameworks were retained as novel contributions.

\mypar{Derive dimensions}
Once coding was complete and all themes had been identified, the first three authors sat together to cluster related themes into higher-level dimensions, capturing broader facets of the data.

\mypar{Perform member checking}
To enhance validity and transparency, we performed member checking~\citep{birt2016member} by sharing the evolving set of codes, descriptions, and linked participant quotes via email. Participants were invited to review and suggest corrections. Three participants expressed agreement with the coding, and one requested a change to a code description, which we incorporated.

\mypar{Analyse and report}
The final set of concepts, themes, and dimensions was analysed and reported in alignment with participant perspectives.
\newcommand{\themePar}[1]{\medskip\noindent\textit{#1}.\ }

\section{Results}
\label{sec:results}

We present the main results answering our research questions. We follow the Gioia methodology, with \dimension{dimensions} highlighted in dark violet boxes and \theme{themes} in light blue. Each theme indicates whether it applies to \textbf{Explainers}, \textbf{Explainees}, or \textbf{Both} perspectives. Themes exclusive to one perspective are marked accordingly.

\subsection{Explanation Topics}
\label{sec:aspects}

To answer \ref{rq1}, we identify two key dimensions capturing topics during software architecture explanations: system concerns and factors influencing architectural decisions. These range from system contexts and architectural views to design rationales and system evolution implications. \Cref{tab:rq3-1} shows the dimensions and themes around this RQ. In this table and the subsequent tables about the dimensions and themes, column \textsf{\textbf{E}} shows whether any \textit{explainer} is concerned with the corresponding theme, and likewise \textsf{\textbf{e}} for \textit{explainees}.

\begin{table}[h]
\footnotesize\sffamily
    \centering
    \caption{\added{Dimensions and themes revolving around \ref{rq1}.}}
    \label{tab:rq3-1}
\begin{tabular}{llcc}
\toprule
\multicolumn{4}{c}{\textbf{\ref{rq1} Explanation Topics}} \\
\midrule
\textbf{Dimension} & \textbf{Theme} & \textbf{E} & \textbf{e} \\
\midrule
\multirow{6}{*}{System Concerns} & The system's context & \checkmark & \checkmark \\
 & Architectural views & \checkmark & \checkmark \\
 & Quality attributes & \checkmark &  \\
 & Architectural styles/patterns & \checkmark &  \\
 & Implementation & \checkmark & \checkmark \\
 & Semantics & \checkmark & \checkmark \\
\midrule
\multirow{3}{*}{Architectural Decision Factors} & Compliance and conformance & \checkmark &  \\
 & Design implications & \checkmark & \checkmark \\
 & Changes and evolution & \checkmark & \checkmark \\
\bottomrule
\end{tabular}
  \vspace{0.5em}
  
\rmfamily
\textbf{\textsf{E}} -- explainer's concern; \textbf{\textsf{e}} -- explainee's concern.
\end{table}

\subsubsection{\dimension{System Concerns}}

This dimension covers the system's context and architectural composition. While architectural views were commonly discussed, quality attributes and specific architectural styles or patterns received notably less attention.

\themePar{\theme{The system's context}\label{theme:context}}
\textbf{Both:} Explainers ($E_2$, $E_5$) and explainees ($e_1$, $e_5$) agree on the importance of understanding the organisational context, enterprise architecture, and business goals for architectural decisions. As $e_5$ states, \textit{``a good explanation puts the architecture in its business context.''}\\
\textbf{Explainees:} Multiple explainees ($e_5$--$e_8$) highlight that, lacking familiarity, they rely on explanations that tie architectural decisions to the business context. Explainee $e_5$ observes that explainers often omit this linkage, causing consistency issues, a concern echoed by $E_9$.  
This reflects an asymmetry: explainers tend to focus on technical knowledge, assuming domain context is known or less relevant, while explainees require the ``why'' before the ``how'' becomes meaningful~\cite{basili2010linking}.

\themePar{\theme{Architectural views}}
\textbf{Both:} Discussions about behaviour are common among participants ($E_1$, $E_2$, $E_5$, $E_7$, $E_8$, $e_3$, $e_6$). Similarly, discussions on structures were widely raised ($E_1$, $E_3$--$E_5$, $E_7$, $e_1$--$e_3$, $e_5$--$e_8$), specifically on the responsibilities and interactions between components. Both explainers and explainees point out the importance of considering complementary views to comprehend the system: \textit{``You want to decouple [the architecture] in different views.''} ($E_5$)\\
\textbf{Explainers:} Deployment and development views are predominantly mentioned by explainers ($E_1$, $E_2$, $E_3$, $E_9$) and were largely absent from explainee responses. This suggests that explainers often aim for sharing a more complete picture of the system's architecture, while explainees take the structural or conceptual perspective as a first step to build a mental model of the system. The discussion on architectural views also anticipates the importance of diagram usage in explanations (cf. \Cref{theme:diagrams}).

\themePar{\theme{Quality attributes}}
\added{Quality attributes remain prevalent in software architecture research \citep{bass2021software,lytra2020quality}. Therefore, we expected them to play a substantial role in architecture explanation practice. However, our findings do not support this.}
\textbf{Explainers:} Attributes such as security, availability, scalability, and flexibility were mentioned only by a few explainers ($E_2$, $E_5$), who discussed how architecture addresses these qualities. For example, $E_2$ mentions: \textit{``We review the quality scenarios and the architectural significant requirements imposed to the software architecture, and we study how those quality scenarios are driven there \ldots\ or how the architecture tries to solve those quality attributes and scenarios.''} \added{This limited mention of quality attributes by the participants (including a total lack of mention by explainees)} indicates under-communication of this topic.  
This carries a risk: Without understanding the rationale for architectural decisions, explainees may miss system limitations or introduce changes that threaten architectural conformance. Future work should address how to integrate quality attributes into explanations while considering explainee's information need.

\themePar{\theme{Architectural styles/patterns}}
\textbf{Explainers:} Different architectural styles, e.g., layered architectures for embedded systems versus microservices, demand tailored explanations ($E_4$, $E_5$). Experienced developers may specifically ask for the rationale behind the choice of style or pattern.  
An identified barrier is that many practitioners lack formal training in architectural terminology, limiting the use of these concepts in explanations ($E_3$, $E_9$). This reinforces the terminology gap between academia and industry and may hinder effective communication with explainees.

\themePar{\theme{Implementation}}
\textbf{Explainers:} \added{Implementation of a system is mainly expressed in the form of source code and other related artefacts.} Discussing implementation details is essential to verify the alignment of the design and the implementation and identify technical debt, especially during code reviews or resolving blockers ($E_1$, $E_3$, $E_7$). As $E_7$ explains: \textit{``\ldots\ when I then have to explain such a feature, then I'm also actively trying to help them \ldots\ both in the sense of where everything should be and what all the parts should be but also how it should look like from an implementation standpoint.''} This characterises a design--implementation drift issue, i.e., ``as-is'' vs. ``as-intended'' architecture.\\
\textbf{Explainees:} Tend to ask code-level questions (e.g., about method purpose) to move their tasks forward ($e_2$, $e_4$).  
Both groups seek actionable understanding, but for different purposes: explainers focus on maintaining architectural integrity; explainees seek to navigate unfamiliar codebases.

\themePar{\theme{Semantics}}
\textbf{Both:} Shared understanding of notations used in an explanation is repeatedly mentioned as essential for effective explanation. \added{These notions range from diagram semantics to domain and architectural terms.} While explicit syntactic questions (e.g., about UML) are rare, participants often discuss how diagram elements (e.g., boxes, arrows) should be interpreted ($E_6$, $E_9$, $e_3$), as $E_9$ describes: \textit{``discussion about the semantics of the word architecture, always [appear] \ldots\ Then the next question is somebody draws an arrow or a box on the white board and it's like, what does that arrow mean? Is it the data flow?''}
Informal or nonstandard diagrams are common, often requiring clarification of meaning ($E_4$, $E_5$, $E_8$, $e_5$, $e_6$). Explainees frequently ask for definitions of domain, business, and architectural terms ($E_3$, $E_9$, $e_3$), and organisational jargon remains a source of confusion ($E_1$, $E_3$, $e_4$, $e_6$).  
Failing to align on semantics can result in faulty mental models and subsequent design or implementation errors.

\subsubsection{\dimension{Architectural Decision Factors}}\label{sec:decision-factors}

Architectural decision factors include requirements, business goals, design implications, and compliance or conformance drivers. These elements help shape architectural choices and are crucial for explanations that align technical decisions with broader organisational needs.

\themePar{\theme{Compliance and conformance}}
\textbf{Explainers:} Explainers consistently stress that architecture must align with both business goals and compliance requirements. These requirements (and consequently alignments) are often poorly documented or implicit ($E_5$--$E_7$). The challenge of making them explicit and traceable is a recurring theme, which often leads to inconsistencies or rework. Technical decisions must also fit within business constraints, such as budget and time ($E_2$, $E_5$, $E_9$), as $E_5$ puts it: \textit{``it's not about only making a nice architecture, but if it [is going to] cost something that is out of budget, \ldots\ it won't work.''} 
Explainers highlight the need to communicate for both technical and business audiences, and to draw from past projects to align with client needs ($E_9$), as well as clarify how features fit the overall architecture ($E_7$). Internal guidelines, like process requirements and documentation protocols, are another focus ($E_6$), indicating attention to organisational compliance even when not formally labelled as such.
Several participants ($E_4$--$E_6$) point to the importance of documenting rationale, linking requirements to architectural decisions, and managing technical debt. 
External compliance with laws, regulations, or contracts further constrains design, but explainers acknowledge the challenge of reconciling business decisions with these policies ($E_2$). In general, explainers act as translators, interpreting often hidden or shifting constraints and ensuring traceability from context to design. Compliance needs may also limit not just the design space but also how architecture can be discussed.

\themePar{\theme{Impacts of decisions}}
\textbf{Explainers:} Understanding past architectural decisions, especially when documentation is lacking, is considered essential to avoid inconsistencies or inefficiencies ($E_1$, $E_3$, $E_5$--$E_7$). Architecture Decision Records (ADRs) are valued for capturing decision rationale ($E_6$, $E_7$). 
Explainers also emphasise the need to discuss alternatives, trade-offs, and impacts on the system with stakeholders ($E_1$, $E_2$, $E_5$), echoed by explainee $e_3$, something that documentation practice recently started embracing via ADRs. $E_8$ highlights the utility of tools for visualising the impact of design choices. However, not every decision warrants detailed documentation; $E_6$ notes that explainers often exercise judgment about what is relevant to share. \\
\textbf{Explainees:} Transparency about the reasoning behind key design decisions is critical for understanding system intent ($e_5$, $e_6$, $e_7$). Explainee $e_5$ specifically expects to see \textit{``key decisions \ldots\ and how they relate to the business goals.''}  
This theme underscores the importance of visible rationale in enabling comprehension, not only for future maintainers but also for new team members.

\themePar{\theme{Changes and evolution}}
\textbf{Explainers:} A core part of the explainer role is to guide explainees in identifying which components can or should be modified, ensuring that evolution maintains conformance and does not lead to architectural drift or erosion ($E_8$, $E_9$). Explainers thus act as gatekeepers, helping others navigate system change responsibly.\\
\textbf{Explainees:} Tracing the history of changes is essential for understanding current implementations, particularly when documentation lags behind the system ($e_3$). Explanation becomes especially critical when changes are not properly recorded, as this can cause confusion during ongoing evolution.\\

\rqanswer{rq1}{
Software architecture explanations typically orbit around the system's context and views, with supporting detail on implementation, semantics, and styles or patterns. They also extend to decision factors such as compliance, rationale, and system evolution. In practice, explainers act as translators: linking design to business goals, constraints, and past choices, while guiding responsible change. Notably, quality attributes, though central to architecture, remain under-communicated.
}













\subsection{Explanation Plan}

\label{sec:plan}
A deliberate plan enhances the effectiveness of software architecture explanations. To answer \ref{rq2}, we structure the plan into three dimensions: activities (what to do), comprehension model (how understanding develops), and explanation style (how it is delivered). These guide a successful explanation. \Cref{tab:rq3-2} shows the dimensions and themes around this RQ.

\begin{table}[h]
\footnotesize\sffamily
    \centering
    \caption{\added{Dimensions and themes revolving around \ref{rq2}.}}
    \label{tab:rq3-2}
\begin{tabular}{llcc}
\toprule
\multicolumn{4}{c}{\textbf{\ref{rq2} Explanation Plan}} \\
	\midrule
\textbf{Dimension} & \textbf{Theme} & \textbf{E} & \textbf{e} \\
\midrule
	\multirow{10}{*}{Activities} & Gain prior knowledge & \checkmark & \checkmark \\
	 & Prepare explanation meeting & \checkmark & \checkmark \\
	 & Review documentation & \checkmark & \checkmark \\
	 & Explore source code & \checkmark & \checkmark \\
	 & Interact with the system & \checkmark & \checkmark \\
	 & Visualise system architecture & \checkmark & \checkmark \\
	 & Provide/receive feedback & \checkmark & \checkmark \\
	 & Support Q\&A & \checkmark & \checkmark \\
	 & Derive working plan & \checkmark & \checkmark \\
	 & Document the outcomes & \checkmark & \checkmark \\
	\midrule
	\multirow{4}{*}{Comprehension Model} & Top-down model & \checkmark & \checkmark \\
	 & Bottom-up model &  & \checkmark \\
	 & Opportunistic model & \checkmark & \checkmark \\
	 & Hybrid model &  & \checkmark \\
	\midrule
	\multirow{4}{*}{Explanation Style} & Incremental explanation & \checkmark & \checkmark \\
	 & Interactive explanation & \checkmark & \checkmark \\
	 & In-situ explanation & \checkmark & \checkmark \\
	 & Non in-situ explanation &  & \checkmark \\
\bottomrule
\end{tabular}
\end{table}

\subsubsection{\dimension{Activities}}
Participants from both perspectives identified a set of core explanation activities, but reported no uniform sequence. We enumerate these activities without suggesting a fixed order, as execution varies by situation, role, and personal preference.

\themePar{\theme{Gain prior knowledge}}
\textbf{Explainees:} Explainees emphasise the need for some level of prior knowledge, whether general software engineering ($e_2$) or system- or domain-specific ($e_7$). However, opinions are split: some ($e_1$, $e_8$) argue domain knowledge is essential to grasp architecture, while others ($e_2$, $e_7$) think that good architectural explanations can bridge knowledge gaps for those with technical backgrounds. Explainee $e_2$ states that, with an effective explanation, \textit{``most people with a software background should be able to understand,''} without needing domain knowledge. \\
\textbf{Explainers:} $E_5$ agrees domain knowledge supports understanding, but also signals the importance of tailoring explanations to the explainee's background. This highlights the need for adaptive strategies and aligning frames of reference, as not all explainees approach explanations with the same knowledge or needs.

\themePar{\theme{Prepare explanation meeting}}
While we did not restrict the format of architecture explanations, nearly all participants recalled dedicated \textit{explanation meetings}, a face-to-face meeting between explainer and explainee with architecture explanation as the agenda.
Both explainers and explainees described that these meetings often require preparation on both sides.  \\
\textbf{Explainers:} Many explainers expect explainees to review background materials and prepare relevant questions in advance ($E_1$, $E_5$, $E_9$). $E_1$ describes sometimes sending explainees away to research and return with questions or proposals: \textit{``Go back and do your research and come back again with the current design and where you want to go.''}  \\
\textbf{Explainees:} Explainees $e_4$ and $e_6$ echo the need for preparation as a shared responsibility: explainees should formulate questions, and explainers should provide clear and structured explanations.

\themePar{\theme{Review documentation}}
\textbf{Both:} Documentation review is a recurring activity, used at various points before or after live explanations ($E_4$, $e_5$, $e_7$).  
Documentation spans architecture diagrams, configuration details, use cases, results from analysis tools, and rationale for key design choices. Explainees often rely on documentation as a first point of contact ($E_8$, $e_2$), or when system experts are unavailable ($e_1$). It also serves as a reference during later development ($e_4$).  
However, most participants recognise documentation is rarely complete or fully aligned with the system, reinforcing the need for supplementary explanation and up-to-date rationale.

\themePar{\theme{Explore source code}}
\label{theme:explore-source-code}
\textbf{Explainers:} Several ($E_3$, $E_4$, $E_7$) use source code exploration, often in pair programming activity, to reinforce architectural concepts and verify alignment with design.  \\
\textbf{Explainees:} Perspectives diverge: some ($e_8$) find code walkthroughs distracting during explanations, saying, \textit{``I would rather have a more overview explanation of the whole system and then [look] myself into the very specific implementations \ldots\ rather than someone telling me the very specific implementations and having to figure out the domain and the overview of the system;''} others ($e_2$) see code review as vital to reinforcing understanding. This split reflects different comprehension models: Some prefer a conceptual foundation first, while others want hands-on implementation details early.

\themePar{\theme{Interact with the system}}
\textbf{Both:} Direct interaction through demos or user exploration helps bridge the gap between abstract architecture and concrete functionality ($E_4$, $e_2$, $e_7$, $e_8$).  
As $e_8$ puts it: \textit{``When I can interact with the system it helps with the explanation. What does a button do? What does this field do? It's easier to follow the explanation.''} Demos reveal system behaviour, clarify feature purpose, and sometimes expose gaps between documentation and reality. ($E_4$: \textit{``Part of that first explanation is showing them or giving them a demo: This is what we built. This is our software. This is where you're going to be living.''})

\themePar{\theme{Visualise system architecture}}
\textbf{Both:} Diagrams and drawings are widely seen as essential for understanding, more effective than text or code alone ($E_1$, $E_4$, $E_6$, $E_8$, $e_5$, $e_6$, $e_7$, $e_8$).  
Visuals clarify structure, foster peer discussion, and make complex interactions tractable. $E_1$: \textit{``What [is] helpful for me to explain my proposal to others is to create diagrams. I think those are the best way to to interact with the fellow architects and other engineers.''}  
Explainees ($e_8$) often describe visualisations as the main explanation aid. (See \Cref{sec:artefacts} for more information on visual artefact types.)

\themePar{\theme{Provide/receive feedback}}
\textbf{Explainers:} Timely feedback from explainees helps clarify ambiguities and improve future sessions ($E_3$). When presenting to external customers, architects expect pushback and use feedback to assess alignment with requirements and compliance ($E_5$, $E_6$), as noted by $E_5$, \textit{``We will go into the architecture presentation and it's just about feedback in our case. Basically, we're proposing something and of course it's not gonna be perfect. Of course, it's gonna be pushed back. I am talking about the good one, good feedback. Hey, you missed this one or why do we do this this way or no; based on our compliance rules, we cannot do that.''}  \\
\textbf{Explainees:} $e_3$ stresses feedback as a tool for clarifying ambiguous points and shaping explanations, shifting from passive reception to active construction of understanding.

\themePar{\theme{Support Q\&A}}
\textbf{Explainers:} About half ($E_1$, $E_3$, $E_8$, $E_9$) stress that supporting Q\&A sessions is essential not just to clarify details, but to examine the understanding of the explainee and to uncover new perspectives. $E_3$: \textit{``You need to give more space for [explainees] to ask questions. Actually, I would say that is one of the most important things.''} $E_9$ notes that recurring questions from past clients help reinforce or adjust explanations using prior insights.\\
\textbf{Explainees:} All explainees ($e_1$--$e_8$) see Q\&A as crucial, both to address knowledge gaps and to follow up incomplete documentation. Several recommend preparing for explanations by reviewing documentation and formulating questions ($e_2$--$e_4$). When documentation is lacking, peer consultation becomes even more important ($e_1$--$e_3$).

\themePar{\theme{Derive working plan}}
\textbf{Both:} Before concluding an explanation, both sides often agree on concrete next steps, like running or debugging the system ($E_1$, $E_5$), carrying out a pilot task ($E_3$, $e_2$, $e_3$), or pair programming ($E_3$).  
$E_3$ describes a hybrid approach: new hires start with pair programming but are encouraged to work independently before seeking help. Later, the team evaluates requirements to define future features ($E_2$, $E_6$, $E_9$). This transition marks the shift from guided explanation to autonomous work.

\themePar{\theme{Document the outcomes}}
\textbf{Both:} Documenting key takeaways from explanations (notes or minutes) ensures a shared understanding of decisions and discussions ($E_6$, $e_7$). As $E_7$ notes: \textit{``I take notes [of] what things are important for what I'm interested in.''}  
Documentation supports explainees in subsequent tasks and helps explainers improve future sessions. In addition, explanation sessions often expose outdated or missing documentation, prompting valuable updates to architectural records.

\subsubsection{\dimension{Comprehension Model}} \label{comprehension-model}
The sequence and nature of explanation activities are shaped by the comprehension model in use. We define the comprehension model as a cognitive framework describing how an individual forms an abstract representation of what is being understood~\cite{storey2005theories}. We build on cognitive models from prior work~\cite{alsaiyd2017source, storey2005theories}: top-down, bottom-up, opportunistic, and hybrid, mapping our themes to this literature.

\themePar{\theme{Top-down model}}
In this model, architecture is understood by linking high-level abstractions down to code as needed~\cite{brooks1983towards}. \\
\textbf{Explainees:} Explainees $e_5$, $e_7$, and $e_8$ endorse this model: $e_8$ explains, \textit{``it is easier to know first what the system is doing as a whole and then focus on different parts of functionality.''}\\
\textbf{Both:} Participants emphasise avoiding information overload. $E_1$: \textit{``I want to limit the information \ldots\ the more information I provide, the harder it will be for the scope and problem to be understood.''} Some, especially in management roles ($E_1$, $E_4$, $E_6$, $E_7$), stay at a high level, focussing on design layout and interactions without deep dives. Explanations often end at the interfaces of key components or services ($E_1$, $E_4$, $E_5$, $e_2$, $e_6$). This points to the need to calibrate abstraction and tailor explanations to the audience.

\themePar{\theme{Bottom-up model}}
This approach starts with inspecting code and low-level details, then builds higher-level models~\cite{brooks1983towards}. \\\textbf{Explainees:} Only $e_1$ explicitly prefers a bottom-up start, usually as part of a combination strategy, beginning with code, then constructing a conceptual model and checking it with the explainer.

\themePar{\theme{Opportunistic model}}
\textbf{Both:} Here, professionals attend to what is most relevant for the immediate task, e.g., refactoring, bug-fixing, or feature work~\cite{alsaiyd2017source, storey2005theories} ($E_7$, $e_7$). Explanation is not for holistic understanding, but for supporting a specific goal~\cite{letovsky1987cognitive, littman1987mental}. $E_7$: \textit{``What I'm considering when explaining things is that [it is] relevant for the tasks that they're doing.''} Focus may shift between architectural views (\Cref{sec:aspects}) ($E_5$, $e_6$, $e_7$), or across abstraction levels ($e_2$). Some ($E_3$) describe actively jumping between perspectives to refine their mental model. This underlines the need for explanations that are responsive to evolving goals and contexts.

\themePar{\theme{Hybrid model}}
This blends the above strategies, switching as needed for the situation~\cite{alsaiyd2017source}. Flexibility allows cross-checking and fills knowledge gaps.\\
\textbf{Explainees:} $e_1$, for example, rarely follows a linear path: starting with the system’s purpose, they alternate between structure (directories, classes) and detailed code, iteratively refining understanding: \textit{``The starting point should be \ldots\ what the system tries to do, then \ldots\ the hierarchy or directories, and then \ldots\ go to each directory \ldots\ and explain what the classes here are doing or how they communicate with each other.''} This iterative, integrative style appears common and pragmatic in handling software complexity.

\subsubsection{\dimension{Explanation Style}}
\label{sec:style}
\textit{Explanation style} concerns how architectural knowledge is conveyed, e.g., its pacing, structure, and supporting resources, distinct from interpersonal communication style. Nevertheless, participants consistently emphasise open communication ($E_1$, $E_2$), minimising conflict ($E_2$), and prioritising clarity and simplicity ($E_5$, $E_7$, $e_4$, $e_6$, $e_8$), aligning with assertive communication principles~\cite{norton2006assertiveness}.

\themePar{\theme{Incremental explanation}}
\textbf{Both:} Both explainers and explainees emphasise the need to spread explanations over multiple sessions, allowing for stepwise knowledge internalisation. $e_8$ stresses the difficulty of absorbing everything at once: \textit{``It's better [if there are] different sessions, so I can understand the information \ldots\ I need time to process a little information and then move on to the next part.''} $E_5$ notes that architectural explanation is rarely ``one meeting,'' but instead, an iterative, interactive process. Rapid, continuous explanations from experts can overwhelm newcomers.

\themePar{\theme{Interactive explanation}}
\textbf{Explainers:} Explainers ($E_1$, $E_8$) structure sessions as interactive, not just presenting but continuously inviting questions and updating diagrams or artefacts. This enables immediate feedback and real-time adjustment.  \\
\textbf{Both:} Many ($E_2$, $e_2$, $e_4$) note that discussions, often with several architects or teammates, are critical for understanding complex issues. $e_2$ highlights the value of ``on-demand'' clarification via chat or informal discussions, reinforcing that accessibility and availability of team expertise are essential for effective architecture comprehension.

\themePar{\theme{In-situ explanation}}
\textbf{Both:} Face-to-face meetings allow explainers to monitor explainees' reactions for signs of confusion or understanding ($E_1$, $E_8$). $E_1$: \textit{``It helps a lot \ldots\ to be face to face with that person and to get feedback on their reaction, because most of the time when I have to explain these terms, I see the other person confused.''} Such in-situ feedback loops drive more adaptive, transparent explanations.

\themePar{\theme{Ex-situ explanation}}
\textbf{Explainees:} When direct sessions are unavailable, explainees such as $e_1$ rely on documents or diagrams in the knowledge base, interpreting them without guidance. This approach is more effortful and prone to misunderstanding, and $e_1$ expresses a preference for live, interactive sessions: \textit{``Having live sessions \ldots\ would be much better to have an easier understanding on how things work.''} This underscores the need for more robust asynchronous or mediated alternatives when face-to-face explanation is not feasible.\\

\rqanswer{rq2}{
An effective architecture explanation plan weaves together three elements: activities, comprehension models, and explanation styles. Activities cover the mechanics: preparation, documentation, system exploration, visualisation, Q\&A, and outcome capture. Comprehension follows different models: top-down, bottom-up, opportunistic, or hybrid, with the top-down model dominating participant's preferences. Explanation styles: incremental, interactive, in-situ or otherwise, govern delivery and pacing. What emerges is not a fixed recipe but a flexible approach: explanations succeed when adapted to audience background, task at hand, and desired depth, ensuring both shared understanding and actionable next steps.
}

\subsection{Supporting Artefacts}
\label{sec:artefacts}
Supporting artefacts are essential for clarifying and improving understanding in software architecture explanations. These include documentation, diagrams aligning with the 4+1 view model, and tools for their creation and management. \Cref{tab:rq3-3} shows the dimensions and themes around \ref{rq3}.

\begin{table}[h]
\footnotesize\sffamily
    \centering
    \caption{\added{Dimensions and themes revolving around \ref{rq3}.}}
    \label{tab:rq3-3}
\begin{tabular}{llcc}
\toprule
\multicolumn{4}{c}{\textbf{\ref{rq3} Supporting Artefacts}} \\
	\midrule
\textbf{Dimension} & \textbf{Theme} & \textbf{E} & \textbf{e} \\
\midrule
	\multirow{2}{*}{Documentation} & Architecture-related documentation & \checkmark & \checkmark \\
	 & Implementation-related documentation & \checkmark & \checkmark \\
	\midrule
	\multirow{6}{*}{Diagrams} & Functional diagrams & \checkmark & \checkmark \\
	 & Process diagrams & \checkmark & \checkmark \\
	 & Development diagrams & \checkmark & \checkmark \\
	 & Deployment diagrams & \checkmark & \checkmark \\
	 & Data diagrams & \checkmark & \checkmark \\
	 & High-level diagrams & \checkmark & \checkmark \\
	\midrule
	\multirow{3}{*}{Tools} & Visual tools & \checkmark & \checkmark \\
	 & Documentation tools & \checkmark & \checkmark \\
	 & Development tools & \checkmark &  \\
\bottomrule
\end{tabular}
\end{table}

\subsubsection{\dimension{Documentation}}
\label{dim:documentation}
Documentation supports explanations both during sessions and independently, especially when explainers are unavailable. Participants describe a wide range of documents, which we group as architecture- and implementation-related.

\themePar{\theme{Architecture-related documentation}}
\textbf{Both:} Beyond diagrams, textual documentation is highly valued ($E_4$, $E_6$, $E_9$, $e_2$, $e_5$, $e_6$, $e_7$). Participants point out that architecture documents are often in unstructured, narrative (``storytelling'') formats: \textit{``Every time we have someone onboarding, we \ldots\ explain all these decisions \ldots\ It's more of a storytelling than [structured documentation], but [it's] is useful for newcomers''} ($E_4$). This embedded context helps new team members grasp rationale and history. By contrast, ADRs provide a structured alternative, capturing decisions, context, alternatives, and stakeholders ($E_5$, $E_6$).

\themePar{\theme{Implementation-related documentation}}
\textbf{Both:} Implementation documentations, like API specs ($E_7$, $e_1$, $e_4$), code documentation ($E_5$, $E_6$, $e_3$, $e_4$), integration manuals ($E_6$), service interaction docs ($E_4$), help explainees understand the system and enable independent exploration and debugging across diverse tech stacks. $E_5$ notes: \textit{``The most important documentation is how to run the system \ldots\ [to] debug the system is the best way of learning it.''} Even low-level documents serve as a bridge for understanding architecture and design, though their completeness and consistency remain a recurring issue.

\subsubsection{\dimension{Diagrams}}
\label{theme:diagrams}
Diagrams are central to explanation, typically aligning with the 4+1 view model~\cite{kruchten19954+1}, but also reflecting high-level and data-centric perspectives. \added{Participants discussed both diagrams that are persisted within documentation or modelling tools, and those that are drawn ad-hoc, e.g., on a whiteboard, during explanations.}

\themePar{\theme{Functional structure diagrams}}
\textbf{Both:} Functional structure diagrams, such as UML class diagrams ($E_1$, $E_2$, $E_8$, $E_9$, $e_2$--$e_4$), component diagrams ($E_1$--$E_3$, $E_6$, $E_8$, $e_3$, $e_4$, $e_8$), dependency diagrams ($e_2$), and layered architecture diagrams ($E_3$), are mentioned frequently.
We surmise that this category of diagrams is particularly valuable at the outset of architectural explanations, as they provide a high-level overview of the system's main components and their interactions.

\themePar{\theme{Process diagrams}}
\textbf{Both:} Process diagrams are almost as popular as structure view diagrams, suggesting that both structural and behavioural aspects are expected to facilitate comprehension.
These are mostly UML sequence diagrams ($E_1$--$E_3$, $E_6$, $E_9$, $e_2$--$e_5$, $e_7$), but there are also references to activity diagrams ($e_2$) and state transition diagrams ($E_6$, $E_8$), as well as informal process diagrams ($E_8$).
State diagrams are mentioned especially for embedded systems, whose behaviour often depends on the system's mode.
This indicates that the usefulness of a type of diagram is context-dependent, i.e., they need to align with the specific characteristics of the system under study.

\themePar{\theme{Development diagrams}}
\textbf{Both:} Development view diagrams (e.g., package diagrams, folder structures) are rarely mentioned ($E_9$, $e_5$, $e_7$), and appear underutilised or replaced by direct code inspection.
This relative absence may suggest that development diagrams are perceived as irrelevant in architectural explanations, with a preference for concrete artefacts (i.e., direct source file access) over abstractions.

\themePar{\theme{Deployment diagrams}}
\textbf{Explainers:} Many explainers use deployment diagrams, network diagrams, infrastructure diagrams, and platform-specific visualisations such as Kubernetes diagrams ($E_2$, $E_3$, $E_5$, $E_6$) to illustrate physical view and how functional components map to the runtime context.  \\
\textbf{Explainees:} Explainees are less likely to note deployment diagrams as useful early on ($e_5$), focussing first on structure and process. Their value seems to grow as the explainee's mental model matures or as specific operational concerns arise.

\themePar{\theme{Data diagrams}}
\textbf{Explainers:} Two explainers ($E_4$, $E_5$) mention the use of entity-relationship diagrams for clarifying data flow.\\
\textbf{Explainees:} One data analytics specialist ($e_1$) highlights the need for dedicated ``data views'' to comprehend data flow and bridge domain concepts and technical artefacts. 
This may become increasingly relevant in response to the increasingly prevalent AI- and data-driven systems~\citep{bosch2021engineering}.

\themePar{\theme{High-level diagrams}}
\textbf{Both:} Participants mention \textit{context diagram}, \textit{use case diagram}, \textit{conceptual diagram}, and \textit{high-level system diagram}, which we consider as \textit{high-level diagrams}: those that portray the system at a high level of abstraction, regardless of the viewpoint ($E_1$, $E_3$--$E_5$, $E_8$, $e_3$, $e_5$, $e_7$).
Generally, participants agree that the complexity of software systems necessitates high-level visual representations to aid explanation and understanding.
Thus, these diagrams appear to serve as entry points into architectural explanation, offering a scaffold on which more detailed system models can later be built.

\subsubsection{\dimension{Tools}}

A wide range of tools support architecture explanation: diagramming, documentation, and development platforms.

\themePar{\theme{Visual tools}}
\textbf{Explainers:} Tools like Draw.io ($E_5$), PlantUML ($E_1$, $E_6$), and GraphViz ($E_9$) are valued for rapid diagram creation, either via visual manipulation, written as code, or generated from other artefacts.  
\textbf{Both:} Some participants rely on custom tools for bespoke needs (e.g., $e_7$, $e_8$, $E_8$), reflecting dissatisfaction with off-the-shelf solutions, although policy constraints may limit adoption ($e_3$). $E_3$ mentions a dedicated microservice observability tool.

\themePar{\theme{Documentation tools}}
\textbf{Both:} Confluence is a common choice for collaborative documentation ($E_8$, $E_3$, $e_7$), with Microsoft Word and PowerPoint used for writing and presenting textual/visual artefacts ($e_4$, $e_5$).

\themePar{\theme{Development tools}}
\textbf{Explainers:} A few explainers report using IDEs ($E_6$, $E_7$) to clarify code-level details, though architecture explanations generally avoid deep dives into source. Interestingly, IDE usage is more often cited by explainers, challenging the expectation that explainers tend to stay at a high level of abstraction.\\

\rqanswer{rq3}{
Effective software architecture explanations rely on diverse artefacts and tools. Documentation spans architecture- and implementation-focused materials, including decision rationales and API specifications. Diagrams aligned with the 4+1 model play a central role, with functional and behaviour views most common. Data views may gain importance; high-level diagram hints the need to manage system complexity. Tools facilitating dynamic diagram creation and collaboration enhance explanations. Occasional use of custom visualisation and development tools reflects varied needs and constraints.
}

\subsection{Typical Questions}
\label{sec:questions}
Q\&A support is a key part of software architecture explanations (\Cref{sec:plan}). Questions relate to topics from \Cref{sec:aspects} but vary in purpose and abstraction. \Cref{tab:rq3-4} shows the dimensions and themes around \ref{rq4}.

\begin{table}[h]
\footnotesize\sffamily
    \centering
    \caption{\added{Dimensions and themes revolving around \ref{rq4}.}}
    \label{tab:rq3-4}
\begin{tabular}{llcc}
\toprule
\multicolumn{4}{c}{\textbf{\ref{rq4} Typical Questions}} \\
	\midrule
\textbf{Dimension} & \textbf{Theme} & \textbf{E} & \textbf{e} \\
\midrule
	\multirow{2}{*}{Question Purposes} & Exploratory questions & \checkmark & \checkmark \\
	 & Confirmatory questions & \checkmark & \checkmark \\
	\midrule
	\multirow{2}{*}{Question Dimensions} & Level of abstraction & \checkmark & \checkmark \\
	 & Architectural viewpoint & \checkmark & \checkmark \\
\bottomrule
\end{tabular}
\end{table}

\subsubsection{\dimension{Question Purposes}}
This dimension explores the purposes underlying questions posed in software architecture explanations.

\themePar{\theme{Exploratory questions}}
\textbf{Both:} Exploratory questions seek to uncover system structure, behaviour, and rationale---asked when the explainee \textit{does not know} and \textit{wants to know}. They typically begin with ``what'', ``how'', or ``why''. As $e_6$ notes: \textit{``[I need] to have [a] full overview of what's happening, why it's happening, how it's happening, choices made. I need to understand the full concepts and \ldots\ why, how, what, when.''}

Two notable subtypes emerge:
\textit{(1) Decision-related questions} probe the rationale for design choices ($E_2$, $E_4$, $E_5$, $e_5$, $e_6$). $E_4$ emphasises: \textit{``The first question is always, `Why?' `Why was this done this way?' That's the main question.''} Such questions elicit detailed reasoning and often reveal inconsistencies between the implemented system and its intended or documented architecture. They may also trigger discussions on the need for refactoring ($E_1$, $E_6$).\\
\textit{(2) Impact questions} examine hypothetical changes and their consequences ($E_9$, $e_7$), e.g., \textit{``What if I change the API?''} or \textit{``What if we reduce the RAM \ldots\ from 192\,GB to 128\,GB?''} Addressing such queries requires deep understanding of how components function and interconnect, and how changes propagate through the system ($e_8$). As $e_8$ notes: \textit{``If you have a what-if question, then you are already close to the answer; if you have that question, you are on the right direction.''}

\themePar{\theme{Confirmatory questions}}
\textbf{Both:} Confirmatory questions are asked when the explainee wants to \textit{confirm what they (think they) know}. They are used to validate or clarify existing understanding, typically yes/no or inductive, and often based on artefacts like diagrams. This helps to deepen baseline comprehension and uncover omissions or errors ($E_6$, $e_7$). For example, $e_7$ describes: \textit{``For instance, one block is connected to five other blocks. You inductively see it, but the person explains only three; then the question [is], `Is it only these three or are there more?' Sometimes there are no more \ldots\ there might be a reason for that, or it might be an error.''}

\subsubsection{\dimension{Question Dimensions}}

Questions during software architecture explanations vary in their level of abstraction and architectural viewpoint. This dimension illustrates how participants navigate between high-level system views and detailed technical inquiries.

\themePar{\theme{Level of abstraction}}
\textbf{Explainers:} Explainers frequently receive detailed questions about modules or components: their responsibilities, behaviours, and roles ($E_1$, $E_3$, $E_5$). $E_1$ observes that queries about component behaviour are common; $E_3$ notes that detail-oriented questions often come from those previously involved with the system. This suggests explainees often seek operational detail, especially in preparation for hands-on engineering tasks.
\textbf{Explainees:} Explainees, however, ask questions across the abstraction spectrum. High-level questions focus on the overall system architecture and its main elements ($e_2$, $e_8$): as $e_7$ puts it, \textit{``For the very high level, what are the big pieces of the system?''} Yet, explainees also delve into specifics: component responsibilities, interfaces, and data flows ($e_1$, $e_4$, $e_8$). These variations reflect not only knowledge gaps, but differing explanatory needs: high-level questions seek coherence and context, while low-level ones are more action-oriented.

\themePar{\theme{Architectural viewpoint}}
\textbf{Explainers:} Two primary viewpoints dominate: functionality (what components do, their capabilities and behaviours) ($E_1$, $E_5$), and interaction (how components relate and communicate) ($E_1$, $E_3$).
\textbf{Explainees:} Explainees, similarly, seek both: they ask about individual component roles and how these contribute to system goals ($e_4$, $e_8$), as well as about integration points and data exchanges across components ($e_2$, $e_3$, $e_4$, $e_1$). As $e_2$ notes: \textit{``Most [questions are] pretty specific. So what does this class do? Or what does this method do?''}\\
The use of both perspectives suggests that architectural understanding is dynamic and situational, requiring iterative synthesis. Rather than treating functionality and interaction as distinct question types, explainees use both in combination to triangulate meaning and uncover rationale.\\

\rqanswer{rq4}{
Typical architecture questions fall into two purposes: exploratory questions probe the unknown, asking ``what’’, ``how’’, or ``why’’ to uncover structure, rationale, and potential impacts of change, while confirmatory questions validate or clarify partial knowledge, often against diagrams or artefacts. These questions operate across two dimensions: level of abstraction (from high-level system overviews to detailed component behaviour) and architectural viewpoint (functionality versus interaction). Together, they show that architectural questioning is dynamic and situational: explainees navigate between broad context and fine-grained detail, using both exploration and confirmation to build, test, and refine their understanding.
}

\subsection{Expectations and Challenges}
\label{sec:expectations}
Participants shared diverse expectations and challenges concerning both the process and outcomes of architectural explanations. These reveal what participants aim to achieve, how they assess explanation quality, and what difficulties arise. To answer \ref{rq5}, we structure findings across three dimensions: expectations during the explanation, expectations after, and challenges encountered during the process. \Cref{tab:rq3-5} shows the dimensions and themes around this RQ.

\begin{table}[h]
\footnotesize\sffamily
    \centering
    \caption{\added{Dimensions and themes revolving around \ref{rq5}.}}
    \label{tab:rq3-5}
\begin{tabular}{llcc}
\toprule
\multicolumn{4}{c}{\textbf{\ref{rq5} Expectations and Challenges}} \\
	\midrule
\textbf{Dimension} & \textbf{Theme} & \textbf{E} & \textbf{e} \\
\midrule
	\multirow{3}{*}{Expectations During Explanation} & Good system documentation &  & \checkmark \\
	 & Knowledgeable explainer &  & \checkmark \\
	 & Actionable closure & \checkmark &  \\
	\midrule
	\multirow{4}{*}{Expectations After Explanation} & System and organisation knowledge & \checkmark & \checkmark \\
	 & Business--architecture alignment & \checkmark & \checkmark \\
	 & Working independence & \checkmark & \checkmark \\
	 & Improved process and results & \checkmark &  \\
	\midrule
	\multirow{5}{*}{Challenges} & Manage multiple levels of abstraction & \checkmark & \checkmark \\
	 & Relevance to task-at-hand & \checkmark & \checkmark \\
	 & Unknown knowledge & \checkmark &  \\
	 & Outsourced services & \checkmark &  \\
	 & Work ownership & \checkmark & \checkmark \\
\bottomrule
\end{tabular}
\end{table}

\subsubsection{\dimension{Expectations During Explanation}}
Expectations during explanations relate to both the quality of the process and the produced outputs. While explainees focus on explanation quality and supporting documentation, explainers' expectations centre on the closure of the session.

\themePar{\theme{Good system documentation}}
As outlined in \Cref{sec:artefacts}, documentation may be referenced during explanations or used independently.  
\textbf{Explainees:} Explainees consistently expect documentation that is clear, detailed, up-to-date, and directly relevant ($e_2$--$e_4$, $e_6$). The absence, obsolescence, or inconsistency of documentation---particularly for architectural artefacts like ADRs---is frequently cited as a barrier ($E_4$, $e_3$--$e_8$). Despite the recognised value, documentation is often incomplete or outdated due to the manual effort required to maintain it.

\themePar{\theme{Knowledgeable explainer}}
\textbf{Explainees:} Explainee $e_6$ expect explanations to be clear, concise, well-structured, and delivered by explainers with sufficient domain knowledge. This expectation underlines the role of explainer credibility in fostering trust and effective knowledge transfer (\Cref{sec:style}, \Cref{sec:aspects}).

\themePar{\theme{Actionable closure}}
\textbf{Explainers:} Explainers prioritise ending sessions with explicit agreements on next steps or follow-up actions ($E_1$, $E_2$, $E_5$, $E_6$, $E_9$): \textit{``Take a note and some actions that you need to do and, yeah, we wrap up with some agreement''} ($E_1$). This reflects the role of architectural explanations within the broader software development feedback loop.
Further research should assess how these closures affect development cycles.

\subsubsection{\dimension{Expectations After Explanation}}
Participants expect that explanations will deliver tangible outcomes. Four themes recur:

\themePar{\theme{System and organisation knowledge}}
\textbf{Both:} Explanations are expected to provide system and organisational understanding. Most ($E_3$, $E_7$, $E_8$, $e_2$, $e_4$, $e_7$) anticipate gaining a high-level overview or ``big picture''; a few ($e_6$, $e_7$) seek complete and detailed understanding, although this depends on context and complexity. Across the board, understanding system's operational context is a baseline expectation.

\themePar{\theme{Business--architecture alignment}}
\textbf{Both:} Both roles expect explanations to clarify how architectural choices serve specific business needs and constraints ($E_5$, $e_5$). Generic justifications are not considered sufficient: \textit{``It has to be \ldots\ this particular architecture is optimised for modifications of that type in this business context''} ($e_5$). This underscores the importance of contextualisation and the need to bridge problem and solution spaces (see twin peaks model~\cite{nuseibeh2001weaving}).

\themePar{\theme{Working independence}}
\textbf{Explainers:} Explainers ($E_3$, $E_4$, $E_7$) expect explainees to become self-sufficient, beginning with simple tasks post-explanation: \textit{``The first step after explanation is to work on something \ldots\ really simple and [then] build up from that''} ($E_4$).  
\textbf{Explainees:} $e_4$ echoes this expectation, but warns that assumptions of full autonomy may not always be realistic and can result in frustration.

\themePar{\theme{Mutual refinement of understanding}}
\textbf{Explainers:} Some explainers note that explanations also refine their own understanding and can lead to improved architectural solutions: \textit{``It's not only that their internal model aligns to mine, it's also that mine aligns to theirs \ldots\ very often the outcome is actually something that is better than what we went into the discussion with \ldots\ I am becoming a better architect''} ($E_6$). Such alignment fosters more effective collaboration and a shared vision.

\subsubsection{\dimension{Challenges}}
Certain questions and topics routinely complicate architecture explanations, often due to required cross-functional knowledge, organisational silos, or lack of direct control. Five challenges are prominent:

\themePar{\theme{Manage multiple levels of abstraction}}
\textbf{Explainers:} Explainers ($E_2$--$E_4$, $E_9$) must tailor the level of abstraction to the audience: some need a broad overview, others require technical detail for design or quality evaluation.  
\textbf{Explainees:} Explainees ($e_7$, $e_2$) agree that too much detail is overwhelming, while insufficient detail leaves gaps. As $e_2$ puts it: \textit{``[The explanation] should consist of different levels \ldots\ classes, components, packages, and the whole file structure.''} The responsibility is on the explainer to strike the right balance.

\themePar{\theme{Relevance to task-at-hand}}
\textbf{Explainers:} Explainers ($E_7$, $E_1$, $E_8$) emphasise focusing content on the explainee's current tasks, to avoid overload and maintain focus: \textit{``What I'm considering when I'm explaining things is that it's relevant for the tasks that they're doing''} ($E_7$).  
\textbf{Explainees:} $e_5$ prefers explanations connected to concrete activities (e.g., implementing features or refactoring), reinforcing the opportunistic model of comprehension.
Explainers once again bear the responsibility of curating content.

\themePar{\theme{Unknown knowledge}}
\textbf{Explainers:} Explainers ($E_3$) may encounter questions outside their expertise, often due to team turnover or unclear ownership: \textit{``They are really asking something that I don't know. For example, something specific about [an] interfaces, but I may not know why it exists''} ($E_3$). Such gaps require effort to bridge, sometimes involving redirection to other experts.

\themePar{\theme{Outsourced services}}
\textbf{Explainers:} Questions about cloud or other external services are difficult when the relevant systems are beyond the team's control: \textit{``We get questions [about things happening in] the cloud \ldots\ which we don't have under control''} ($E_6$). The organisation often acts as an intermediary, and incomplete knowledge can block full explanation.

\themePar{\theme{Work ownership}}
\textbf{Explainers:} Unclear or missing ownership complicates explanations and maintenance ($E_6$).  
\textbf{Explainees:} Explainees ($e_1$, $e_6$) find it difficult to proceed when responsibility for system parts is ambiguous or monopolised by a single person: \textit{``[You do not want to] become dependent on \ldots\ one person knowing that information, because \ldots\ some things will forever be unanswered.''} ($e_6$). Clear ownership and knowledge sharing are critical for continuity and learning.\\

\rqanswer{rq5}{
Explainees expect clear, structured explanations from knowledgeable explainers, supported by accurate and accessible documentation. They value relevance to their immediate tasks and aim to gain a contextual understanding of the system, its architecture, and its business alignment, ultimately enabling independent work. Explainers, in turn, focus on delivering actionable closure and see explanations as a means to refine their own understanding and architectural decisions. They face challenges: explainers must manage appropriate abstraction levels, curate relevant content, and navigate gaps in knowledge, especially around outsourced services or unclear ownership. Documentation issues further complicate the process for both parties.
}

\color{blue}

\section{Discussion}
\label{sec:discussion}

We now interpret and synthesise our empirical findings by first, presenting software architecture explanation through the Twin Peaks lens~\citep{nuseibeh2001weaving,ward1985structured}; second, introducing the Explanation Window framework as an integrative device to manage the complexity of explaining software architecture; and third, describing implications for practitioners. Throughout, we indicate how empirical themes and participant accounts motivated these theoretical contributions.


\subsection{Explanation Through the Twin Peaks Lens}

\begin{figure*}[h]
    \centering
    \includegraphics{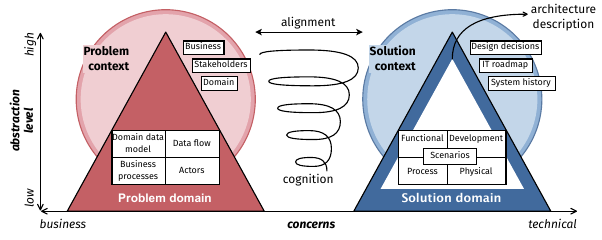}
    \caption{The Twin Peaks model: Software architecture explanation requires explicit alignment of business and technical context, including tacit background (the context ``halos'') on both the problem and solution domains.}
    \label{fig:twin-peaks}
\end{figure*}

The Twin Peaks model, depicted in \Cref{fig:twin-peaks}, describes software development as the co-evolution of requirements (problem domain) and architecture (solution domain), where each peak informs and constrains the other through iterative refinement~\citep{nuseibeh2001weaving}. This framing is particularly relevant for software architecture explanations, which, based on our findings, require articulation of both structural design decisions and the rationale linking them to stakeholder needs and system constraints. Explanations therefore continuously jump across both peaks: they clarify how requirements motivate architectural choices and how architectural possibilities reshape requirements. Hereby, we ground some of our findings on the Twin Peaks model by addressing iterative alignment alongside cognitive and contextual constraints.

\subsubsection{Iterative Alignment}
\label{sec:iterative-alignment}

A central empirical finding is that software architecture explanations operate as processes of \textit{iterative alignment} between the mental models of explainers and explainees. This emerged from themes such as \theme{business--architecture alignment}, \theme{system and organisation knowledge}, and frequent mentions of the need to connect architectural choices with business constraints and stakeholder needs (see \Cref{sec:aspects}). 
This dual alignment is not static: several explainers reported refining their explanations in response to real-time feedback (\theme{interactive explanation}, \theme{provide/receive feedback}). This also reflects a two-way negotiation of understanding (\theme{mutual refinement of understanding}).

The empirical data also reveal a recurring challenge: participants often encountered gaps in contextual knowledge, which we interpret as the ``halos'', representing tacit contextual knowledge on both sides, that we added to the Twin Peaks model (\Cref{fig:twin-peaks}). These halos reflect the often unstated but influential background, such as organisational practices, compliance expectations, or domain conventions, which our participants repeatedly highlighted as shaping explanations. Reflecting our own experience and interpretation in light of our findings, business stakeholders rarely articulate the operational logic behind their needs; developers, in turn, often omit the rationale behind design choices, assuming technical knowledge. These tacit assumptions can quietly shape the explanation. Practitioners must actively surface these contextual dependencies. Otherwise, critical background may be misaligned or omitted in the explanation despite accurate factual transfer.

When explanations fail, they often expose underlying issues with the peaks-alignment, which manifests as undocumented system changes, unclear \theme{work ownership}, or \theme{unknown knowledge}. Thus, these failures act as diagnostic probes, revealing systemic issues. Viewed this way, explanations are not merely pedagogical events but continuous architectural alignment across roles and concerns.

\subsubsection{Cognitive and Contextual Constraints}
\label{sec:cognitive-contextual}

Empirical codes such as \theme{manage multiple levels of abstraction}, \theme{incremental explanation}, and \theme{top-down model} directly support our identification of cognitive constraints. Participants consistently described the challenge of not overwhelming the audience, reporting strategies like starting broad before zooming into details. 
These concerns about information overload and adapting to the explainee's prior knowledge (\theme{gain prior knowledge}) resonate with cognitive load theory~\citep{sweller1988cognitive} and theories of working memory~\citep{oberauer2019working}, showing that explanation must be calibrated to avoid overload. The opportunistic shifts that we observed align with prior accounts of non-linear comprehension in software engineering~\citep{letovsky1987cognitive,littman1987mental}.

Additionally, task context was highlighted: Our initial assumption was that architecture explanation is primarily conceptual and abstract, distinct from task-driven, operational comprehension at the source code level. However, the data shows that explanations are often tightly coupled to the explainee's immediate goals (\theme{relevance to task-at-hand}) and structured accordingly.
Yet, calibration relied almost entirely on the explainer's judgment (e.g., reflecting the preferred comprehension model) instead of structured processes.

\subsection{The Explanation Window}
\label{sec:explanation-window}

Building on the need for iterative alignment (Section~\ref{sec:iterative-alignment}) and the cognitive and contextual constraints (Section~\ref{sec:cognitive-contextual}), we conceptualise the \textit{Explanation Window} as a device to balance scope and depth in software architecture explanations. 

We define the Explanation Window as a \emph{rectangle} through which the system is viewed during an explanation. Its \emph{height} corresponds to the depth of detail (ranging from high-level abstractions down to code-level specifics), while its \emph{width} corresponds to the scope of functionality covered (ranging from a narrow slice of the system to a broad set of features). The window constrains the explanation: widening the scope necessarily reduces the feasible depth, while increasing depth requires narrowing the scope. Explanations remain cognitively manageable only when the ``window area'', i.e., the product of depth and scope, stays within practical limits yet to be defined through further investigation.

\begin{figure}[!ht]
    \centering
    \includegraphics{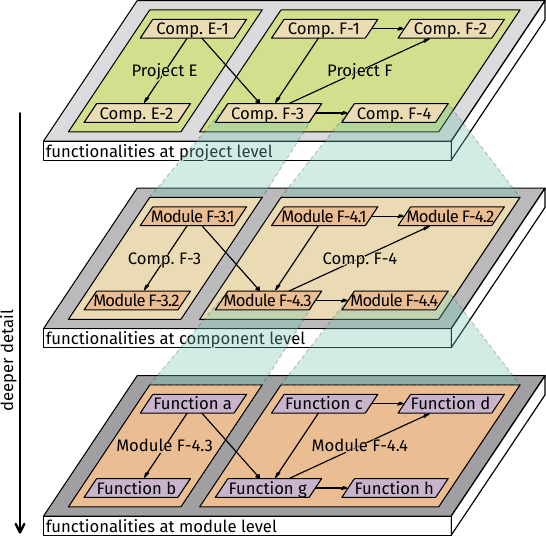}
    \caption{The Explanation Window can be situated on and across different planes (project, component, module, function) representing natural abstraction levels, constraining how much of the system is visible at once.}
    \label{fig:explanation-window}
\end{figure}

\Cref{fig:explanation-window} situates the Explanation Window on three hierarchical planes of abstraction: the \textit{project level} (capturing entire projects and their components), the \textit{component level} (showing modules within components), and the \textit{module level} (breaking down to individual functions). Each plane provides a natural unit of reasoning, but only a bounded portion can be viewed at once. An architect may open a wide but shallow window at the project level to explain overall organisation and dependencies to business stakeholders. In contrast, when explaining to a developer trying to solve a bug, they may open a narrow but deep window at the module level to reason about the behaviour of specific functions, with only a small fraction of the broader architecture in view. This underlines that explanations are inherently selective: the same system can be framed differently depending on role, purpose, and the trade-off between scope and depth.

\begin{figure}[!ht]
    \centering
    \includegraphics{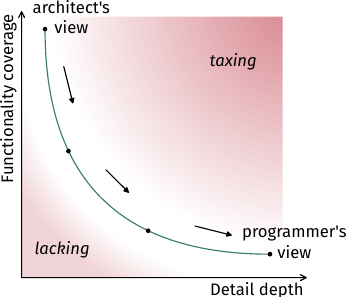}
    \caption{The trade-off between depth and scope: The Explanation Window occupies a position on a curve defined by detail depth ($x$-axis) and functionality coverage ($y$-axis). The effective region corresponds to rectangles whose area remains within cognitive limits.}
    \label{fig:window-plane}
\end{figure}

\Cref{fig:window-plane} frames this trade-off between depth and scope more abstractly. The $x$-axis represents detail depth, the $y$-axis functionality coverage, and the curve bounds the region in which explanations remain effective. Deviating from the curve risks ``lacking'' information or ``taxing'' cognition.

The window also has a temporal dimension: its \textit{trajectory}, i.e., how it moves across depth and scope during an explanation. These trajectories can follow different comprehension models. Top-down approaches begin with wide, shallow windows and gradually roll down into narrower, deeper ones; opportunistic approaches shift dynamically in response to tasks or questions; bottom-up and hybrid approaches, though less frequently observed, represent alternative ways the window can move.

These trajectories are further shaped by variables such as the explainee's role (e.g., programmer vs.\ business stakeholder), familiarity with the project, and immediate task context (e.g., onboarding, debugging, integration). Each combination demands a differently sized and sequenced window: onboarding may require a wide but shallow rectangle; integration may require cross-cutting windows across planes; debugging concentrates on narrow, deep windows. While these examples are interpretative illustrations rather than empirical results, they clarify how the Explanation Window can be operationalised in practice as a flexible, adaptive guide for balancing scope, depth, and task relevance.

\subsection{Implications for Practice}
\label{sec:implications}

Our findings suggest several actionable implications for improving architectural explanations.

\mypar{Plan explanations around scope and abstraction level} Participants' frequent challenges to \theme{manage multiple levels of abstraction} imply the importance of deliberate planning. The interpretative Explanation Window concept directly addresses this empirically-supported need by providing a practical structure for adapting explanations to the explainee's role, task, and comprehension model.

\mypar{Shift explanation towards joint reasoning} Participants explicitly valued interactive Q\&A and feedback mechanisms (\theme{interactive explanation}, \theme{provide/receive feedback}). Explainers adjust their framing based on feedback; explainees challenge gaps or assumptions. This interactivity helps align mental models on both sides, and often improves the explainer's own grasp of the architecture.

\mypar{Ensure actionable closure} Participants clearly expect explanations to lead to concrete actions (\theme{actionable closure}). This not only orients explainees toward \theme{working independence}, but also tests whether understanding was sufficient to proceed. Closure reinforces \theme{relevance to task-at-hand} already implicit in most explanation requests.

\mypar{Prioritise automated documentation} Participants explicitly identify documentation problems as frequent (\theme{good system documentation}). Although participants do not explicitly label manual documentation untenable, repeated mentions of outdated and inconsistent documentation strongly support investment in automated solutions to maintain accuracy. Automated, extractive documentation, e.g., visualisation of current module dependencies, recent changes, or active interfaces, can ensure explanations are grounded in what the system actually is, not what it was.\\
Additionally, we lay the groundwork for knowledge criteria to be used in such automation. 
This study does not aim to define a complete set of criteria for automated explanations. However, it surfaces knowledge dimensions that are necessary (without claiming sufficiency) for this vision: coverage of multiple views, cross-abstraction navigation, explicit rationale, and separation of problem- and solution-domain concepts. 
These provide a foundation for future work to consolidate them into a more comprehensive set of knowledge criteria. 
Such criteria would then guide the design of suitable knowledge representations and support the evaluation of automated architectural explanations.\\

Ultimately, an effective architectural explanation involves balancing comprehensiveness with cognitive manageability.  The Explanation Window provides both a conceptual and a practical lens for doing so. We believe that it can inform tools, practices, and training materials that are sensitive to the dual challenges of technical complexity and human cognition.

\color{black}
\section{Trustworthiness}
\label{sec:ttv}

We assess the trustworthiness using the Total Quality Framework, \added{which is a more suitable approach for qualitative studies than the threats-to-validity lens common in quantitative research, because it addresses quality holistically across the design, conduct, and interpretation of qualitative studies rather than reducing it to a binary validity judgment}~\citep{roller2015applied}. We consider the credibility, analysability, transparency, and usefulness facets.

\subsection{Credibility}
\added{Credibility refers to the completeness and accuracy of the collected data. To strengthen credibility, novice participants were required to articulate their understanding of software architecture before interviews, ensuring suitability for the study and allowing us to contextualise their subsequent responses. The interview guide was piloted and revised to reduce ambiguity and confusion. To ground responses in real-world experience, we explicitly asked participants to reflect on their actual working contexts. Because we relied on a retrospective self-report, the study is subject to recall bias and cannot fully capture ``naturalistic'' behaviour; participants may omit, reinterpret, or selectively recall experiences. We attempted to mitigate this by instructing participants to anchor responses in recent, concrete work situations and by triangulating accounts across multiple participants, but the absence of in situ data collection remains a constraint. Anonymity and data privacy rights were guaranteed to promote honest participation. Member checking was carried out by providing participants with codes derived from their own transcripts; only one participant requested a minor correction.}

\added{Our recruitment combined convenience, purposive, and snowball sampling. These approaches may introduce sample bias and limit diversity across organisations, industries, or cultures. Relying on professional and academic networks likely excluded practitioners from less connected environments. We did not employ paid online platforms such as Prolific for participant recruitment; our rationale is that such platforms rarely provide targeted access to practitioners in specialised software architecture roles, and quality control for professional background is difficult to guarantee. Although these services may suit studies targeting broad end-user populations, they were not suitable for our aim of in-depth, context-sensitive inquiry among experienced professionals. We acknowledge that this decision limits generalisability, and future research should combine sampling methods to address this gap.}

\added{Because key constructs can be interpreted differently in technical interview studies, we explicitly asked all novice or early-career participants to define ``software architecture'' prior to the main interview. This served two purposes: first, it confirmed that participants possessed a relevant baseline understanding; second, it allowed us to interpret their subsequent responses in light of their own conceptualisation. In doing so, we reduce the risk of credibility threats arising from inconsistent or assumed meanings of central terms. This practice makes explicit how participants' perspectives were anchored and how our interpretations accounted for their definitions.}

\subsection{Analysability}
Analysability concerns the completeness and rigour of data analysis and interpretation. We followed the Gioia methodology, using systematic thematic coding with iterative development and refinement of codes. Four researchers participated in coding and interpretation, mitigating individual bias and increasing consistency. Interpretive disagreements were resolved through discussion and consensual decision rules. The analysis was anchored in established frameworks and relevant literature.

\subsection{Transparency}
Transparency is supported by detailed reporting of all methodological decisions, including the rationale for semi-structured interview design, pilot testing, stepwise coding, member checking, and participant selection. Anonymised transcripts and the codebook are publicly accessible for audit and secondary analysis.

\subsection{Usefulness}
Usefulness refers to the practical value and transferability of the findings. Our results provide grounded, practice-oriented insights into the realities of software architecture explanation, potentially relevant to professionals in similar roles or organisational contexts. However, our sampling was not random or exhaustive, so transferability is limited to contexts with similar characteristics. Future research should seek broader and more heterogeneous samples, potentially leveraging online professional communities or mixed-method sampling, to improve representativeness and robustness.

\added{Because interviews were conducted remotely---often outside the immediate physical workspace of participants---our method does not maximise contextual realism as immersive ethnography or in situ participant observation would. We could not independently verify accounts against artefacts, live practices, or organisational context. As such, our findings should be interpreted as authentic perspectives of practitioners, not direct observations of behaviour. This necessarily restricts generalisability; transferability depends on the reader's judgment regarding similarity of their own context to those reported here.}

\section{Conclusions}
\label{sec:conclusions}

This study examined how professionals explain software architecture in practice, using semi-structured interviews with both explainers and explainees across a range of organisations. We found that effective architectural explanation extends beyond factual or technical accuracy: it is shaped by the explainee's immediate goals and role, the explainer's judgment, and the need to align business context with system structure. Rather than being a purely abstract process, architecture explanation is often driven by practical needs, especially those arising from changing requirements, onboarding, and day-to-day operational demands.

Our findings show that explanations necessarily span both the problem and solution domains, integrating stakeholder concerns, business rules, technical constraints, and implementation details. Effective explanations do not simply recapitulate architectural documentation; rather, they assemble and connect information from a wide variety of sources surrounding the software system, such as business context (stakeholder interests, compliance requirements), development history (past decisions, evolution), and system rationale. Explanations frequently make explicit the relationships or traceability between these different pieces of information, offering context and justifications that are rarely consolidated in any single artefact.

Managing the complexity of explanation is a central challenge. Practitioners address this by adapting abstraction levels and scoping content to the explainee's role and goals, often refining their approach interactively in response to feedback. Explanation thus emerges less as one-way transfer than as collaborative, iterative reasoning.

To make sense of these practices, we introduced the Explanation Window as a conceptual tool to balance the scope and depth of explanations, and we related our findings to the Twin Peaks model. The Explanation Window frames how content is selected and presented according to task, role, and context, and its movement over time reflects varying comprehension strategies. These frameworks provide a foundation for structuring, diagnosing, and adapting explanations to manage both cognitive and contextual constraints.

\added{Our study provides a first systematic characterisation of architectural explanation in industry, but we emphasise its exploratory nature. As a judgement study based on self-reported accounts, our findings capture context-rich practitioner perspectives rather than naturalistic observations. Future research should deepen understanding of how factors such as context, role, prior experience, and system familiarity shape explanation practices. Extending this work through in-situ studies, triangulation with artefacts, and broader sampling would strengthen transferability and provide a basis for practical methods and tools. Further work is also needed to develop and evaluate approaches that support the creation and maintenance of dynamic, real-time architectural explanations, for both expert audiences and for onboarding and cross-disciplinary collaboration. While our study surfaces knowledge dimensions necessary for defining criteria to support automation, consolidating these into comprehensive frameworks remains a task for future research. Such efforts may leverage the adapted Twin Peaks model and the Explanation Window to guide information selection and diagram slicing, and explore emerging technologies, such as AI-driven explanation assistants, to enhance the clarity, traceability, and efficiency of software architecture explanations across diverse organisational settings.}

\bibliographystyle{spbasic}
\bibliography{manuscript}

\end{document}